\documentclass[aps,prd,preprint,superscriptaddress,tightenlines,nofootinbib]{revtex4}

\usepackage{graphicx}
\usepackage{dcolumn}
\usepackage{bm}

\usepackage{graphics}
\usepackage{epsfig}
\usepackage{amsmath}

\begin{document}

\preprint{CLNS 06/1956}
\preprint{CLEO 06-05}

\title{Measurement of Interfering $K^{*+}K^{-}$ and $K^{*-}K^{+}$ Amplitudes \\ in the Decay $D^{0} {\rightarrow} K^{+} K^{-} \pi^{0}$}

\author{C.~Cawlfield}
\author{B.~I.~Eisenstein}
\author{I.~Karliner}
\author{D.~Kim}
\author{N.~Lowrey}
\author{P.~Naik}
\author{C.~Sedlack}
\author{M.~Selen}
\author{E.~J.~White}
\author{J.~Wiss}
\affiliation{University of Illinois, Urbana-Champaign, Illinois
61801}
% ...
\collaboration{CLEO Collaboration}
\noaffiliation

\date{June 14, 2006}

\begin{abstract}
We have studied the Cabibbo-suppressed decay mode $D^{0} \rightarrow
K^{+} K^{-} \pi^{0}$ using a Dalitz plot technique and find the
strong phase difference $\delta_{D} \equiv \delta_{K^{*-} K^{+}}
-\delta_{K^{*+} K^{-}} = 332^{\circ} \pm 8^{\circ} \pm 11^{\circ}$
and relative amplitude $r_{D} \equiv
a_{K^{*-}K^{+}}~/~a_{K^{*+}K^{-}} = 0.52 \pm 0.05
 \pm 0.04$. This measurement indicates significant destructive interference
between $D^{0}\rightarrow K^{+}(K^-\pi^0)_{K^{*-}}$ and
$D^{0}\rightarrow K^{-}(K^+\pi^0)_{K^{*+}}$ in the Dalitz plot
region where these two modes overlap. This analysis uses
$9.0~{\operatorname{fb}}^{-1}$ of data collected at $\sqrt{s}
\approx 10.58~\operatorname{GeV}$ with the CLEO III detector.
\end{abstract}

 \maketitle

The determination of the Cabibbo-Kobayashi-Maskawa (CKM) angle
$\gamma$ (also referred to as $\phi_3$) is important, yet
challenging. Currently $\gamma$ is inferred to be
${58.6^\circ}^{+6.8^\circ}_{-5.9^\circ}$ from various experimental
and theoretical constraints \cite{GammaPaper}. Grossman, Ligeti, and
Soffer \cite{Grossman} have proposed a method for a direct
measurement of $\gamma$ by studying $B^\pm \rightarrow D K^\pm$,
where the neutral $D$ meson ($D^0/\overline D{^0}$) decays to
$K^{*+} K^-$ or $K^{*-} K^+$. An important ingredient in this
analysis is the knowledge of the relative complex amplitudes of
$\overline D{^0} \rightarrow K^{*+} K^{-}$ and $D^{0} \rightarrow
K^{*+} K^{-}$, which, in the absence of CP violation, is the same as
that between $D^{0} \rightarrow K^{*-} K^{+}$ and $D^{0} \rightarrow
K^{*+} K^{-}$. The main goal of the analysis described here is to
measure the strong phase difference $\delta_D$ and relative
amplitude $r_D$ between $D^{0} \rightarrow K^{*-} K^{+}$ and $D^{0}
\rightarrow K^{*+} K^{-}$, which is required for the proposed
extraction of $\gamma$. We are further motivated by a recent paper
of Rosner and Suprun \cite{Rosner} that points out the sensitivity
to $\delta_D$ using $D^{0} \rightarrow K^{+} K^{-} \pi^{0}$ produced
in $e^+e^- \rightarrow \psi(3770) \rightarrow D^{0}\overline D{^0}$,
though the analysis presented here relies on $D^{0}$ mesons from
$D^{*+}$ meson decays in $e^+e^-$ continuum production at $\sqrt{s}
\approx 10.58~\operatorname{GeV}$. This is the first analysis of the
resonant substructures of $D^{0} \rightarrow K^{+} K^{-} \pi^{0}$
and their interference. The relevant published individual branching
ratios ($BR$) are $BR(D^{0}\rightarrow K^{+}K^{-}\pi^{0}) = (0.13\pm
0.04)\%$, $BR(D^{0}\rightarrow K^{\ast +}K^{-}) = (0.37\pm 0.08)\%$,
$BR(D^{0}\rightarrow K^{\ast -}K^{+}) = (0.20\pm 0.11)\%$, and
$BR(D^{0}\rightarrow \phi\pi^0) = (0.076\pm 0.005)\%$
\cite{PDG,oldKKpi,oldKstarCLEO,oldKstarARGUS,PhiPi0BELLE}.

Three-body decays of $D$ mesons are expected to be dominated by
resonant two body decays~\cite{bsw,bedaque,chau,terasaki,Buccella}
and the well established Dalitz plot analysis technique
\cite{DalitzPDG} can be used to explore their relative amplitudes
and phases. The CLEO collaboration has published Dalitz plot
analyses for several three-body $D^0$ decays over the past few years
\cite{Tim,A,B,Asner:2003uz,C,D} and the work described here closely
follows the methods developed in these previous analyses.

This analysis uses an integrated luminosity of
$9.0~\operatorname{fb}^{-1}$ of $e^+e^-$ collisions at $\sqrt{s}
\approx 10.58~\operatorname{GeV}$ provided by the Cornell Electron
Storage Ring (CESR). The data were collected with the CLEO III
detector \cite{Kubota:1992ww,Hill:1998ea,CLEOIII}. To suppress
backgrounds and to tag the flavor $D^0 (\overline D{^0})$, the $D^0$
mesons are reconstructed in the decay sequence
$D^{*+}\rightarrow\pi^+_s D^0$, where the sign of the slow pion
$\pi^+_s (\pi^-_s)$ tags the flavor of the $D^0 (\overline D{^0})$
at the time of its production.

The detected charged particle tracks must reconstruct to within 5 cm
of the interaction point along the beam pipe and within 5 mm
perpendicular to the beam pipe (the typical beam spot is
300~${\mu}$m in the horizontal dimension, 100~${\mu}$m in the
vertical dimension, and 10 mm in the longitudinal dimension). The
cosine of the angle between a track and the nominal beam axis must
be between $-0.9$ and $0.9$ in order to assure that the particle is
in the fiducial volume of the detector. The $\pi_s$ candidates are
required to have momenta $150 \le p_{\pi_s} \le
500~\operatorname{MeV}\!/c$, and kaon candidates are required to
have momenta $200 \le p_{K} \le 5000~\operatorname{MeV}\!/c$.
Candidate kaon tracks that have momenta greater than or equal to
$500~\operatorname{MeV}\!/c$ are selected based on information from
the Ring Imaging Cherenkov (RICH) detector \cite{RICH} if at least
four photons associated with the track are detected. The pattern of
the Cherenkov photon hits in the RICH detector is fit to both a kaon
and a pion hypothesis, each with its own likelihood ${\cal{L}}_K$
and ${\cal{L}}_\pi$. We require $(-2
\operatorname{ln}{\cal{L}}_K)-(-2 \operatorname{ln}{\cal{L}}_\pi) <
0$ for a kaon candidate to be accepted. Candidate kaon tracks
without RICH information or with momentum below
$500~\operatorname{MeV}\!/c$ are required to have specific energy
loss in the drift chamber within 2.5 standard deviations of that
expected for a true kaon.

The $\pi^0$ candidates are reconstructed from all pairs of
electromagnetic showers that are not associated with charged tracks.
To reduce the number of fake $\pi^0$s from random shower
combinations, we require that each shower have an energy greater
than $100~\operatorname{MeV}$ and be in the barrel region of the
detector. The two photon invariant mass is required to be within 2.5
standard deviations of the known $\pi^0$ mass. To improve the
resolution on the $\pi^0$ three-momentum, the $\gamma\gamma$
invariant mass is constrained to the known $\pi^0$ mass.

\begin{figure}
\includegraphics*[width=3.410in]{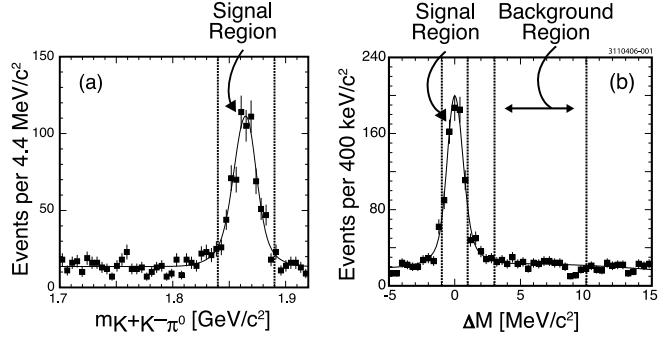}
\caption{Distribution of (a) $m_{{K^+}{K^-}{\pi^0}}$ for
$|{\Delta}M| < 1~\operatorname{MeV}\!/c^2$ and (b) ${\Delta}M$ for
$1.84 < m_{K^+K^-\pi^0} < 1.89~\operatorname{GeV}\!/c^2$ after
passing all other selection criteria discussed in the text. The
solid curves show the results of fits to the $m_{{K^+}{K^-}{\pi^0}}$
and ${\Delta}M$ distributions, respectively. The vertical lines in
(a) and the left-most set of vertical lines in (b) denote the signal
region. The right-most set of vertical lines in figure (b) denote
the ${\Delta}M$ sideband used for estimation of the background
shape.} \label{fig:FigureOne}
\end{figure}

We reconstruct the decay chain $D^{*+}\rightarrow\pi^+_s D^0,~D^{0}
\rightarrow K^{+} K^{-} \pi^{0}$ with the requirement that the
$D^{\ast+}$ momentum be at least as large as one-half of its maximum
allowed value in order to suppress large combinatoric backgrounds
from $B$-meson decays. The $D^0$ candidate invariant mass
$m_{K^+K^-\pi^0}$ and invariant $D^{*+}-D^0$ mass difference $\Delta
M~\equiv~m_{K^{+}K^{-}\pi^{0}\pi^+_s}-m_{K^{+}K^{-}\pi^{0}}-(m_{D^{*+}}-m_{D^0})$
are calculated for each candidate, where $m_{D^{*+}}$ and $m_{D^0}$
are taken from Ref.~\cite{PDG}. The distributions of
$m_{K^+K^-\pi^0}$ and ${\Delta}M$ are shown in
Fig.~\ref{fig:FigureOne}. We fit each of the distributions to the
sum of two bifurcated Gaussians plus a background shape which is
constant (for $m_{K^+K^-\pi^0}$) or parabolic (for ${\Delta}M$). We
find an average signal fraction of $(81.8 \pm 6.3  \pm 2.8 )\%$,
where the systematic error is half of the difference between the
signal fraction from the fits to $m_{K^+K^-\pi^0}$ and ${\Delta}M$.
We select a signal region defined by $1.84 < m_{K^+K^-\pi^0} <
1.89~\operatorname{GeV}\!/c^2$ and $|{\Delta}M| <
1~\operatorname{MeV}\!/c^2$ which contains 735 $D^{0} \rightarrow
K^{+} K^{-} \pi^{0}$ candidates.

We expect CP violation in $D$ decay to be negligible and assume the
amplitudes for $D^{0} \rightarrow K^{*-}K^{+}$ and $D^{0}
\rightarrow K^{*+}K^{-}$ are equal to the amplitudes for
charge-conjugated modes $\overline D{^0} \rightarrow K^{*+}K^{-}$
and $\overline D{^0} \rightarrow K^{*-}K^{+}$, respectively. This
allows us to double our statistics in a single Dalitz plot by
combining flavor-tagged $D^{0} \rightarrow K^{+} K^{-} \pi^{0}$ and
$\overline D{^0} \rightarrow K^{+} K^{-} \pi^{0}$ candidates and
choosing the $m^2_{K^-\pi^0}$ variable for one to be the
$m^2_{K^+\pi^0}$ variable for the other (and vice versa). The
inclusion of charge conjugate modes is implied throughout this
paper.

\begin{figure}
\includegraphics*[width=3.410in]{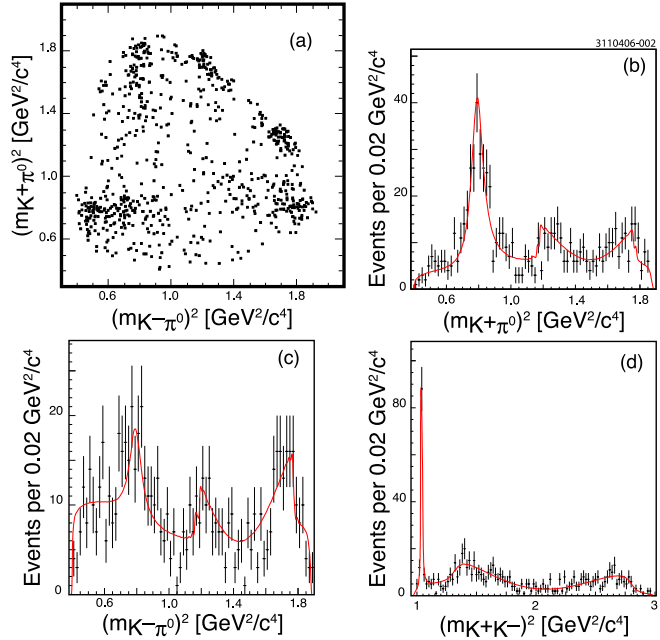}
\caption{(a) The Dalitz plot distribution for $D^{0} \rightarrow
K^{+} K^{-} \pi^{0}$ candidates. (b)-(d) Projections onto the
$m^2_{K^+\pi^0}$, $m^2_{K^-\pi^0}$, and $m^2_{K^+K^-}$ axes of the
results of Fit A showing both the fit (curve) and the binned data
sample. The curves of Fit B projections are indistinguishable from those of Fit A.}
 \label{fig:FigureTwo}
\end{figure}

Figure~\ref{fig:FigureTwo}(a) shows the Dalitz plot distribution for
the $D^{0} \rightarrow K^{+} K^{-} \pi^{0}$ candidates. The enhanced
bands perpendicular to the $m^2_{K^-\pi^0}$ and $m^2_{K^+\pi^0}$
axes at an invariant mass-squared of $m^2_{K\pi} \approx
0.8~\operatorname{GeV^2}\!/c^4$ correspond to $K^{*}(892)^-$ and
$K^{*}(892)^+$ resonances, respectively. The $\phi(1020)$ can be
seen as a diagonal band along the upper right edge of the plot. The
vector nature of these resonances is evident from the depleted
region in the middle of each band. The nearly missing bottom lobe of
the $K^{*}(892)^-$ band and the enhanced left lobe of the
$K^{*}(892)^+$ band show that these resonances are interfering with
opposite phases with the $S$-wave amplitude under these resonances.

We parameterize the $D^{0} \rightarrow K^{+} K^{-} \pi^{0}$ Dalitz
plot following the methodology described in
\cite{Tim,A,B,Asner:2003uz,C,D}. We express the amplitude for $D^0$
to the $j^{\text{th}}$ quasi-two-body state as $a_j
e^{i\delta_j}{\cal B}_j^{(k)}$, where $a_{j}$ is real and positive
and ${\cal B}_j^{(k)}$ is the Breit-Wigner amplitude for resonance
$j$ with spin $k$ described in Ref.~\cite{Tim}. Our sign convention
implies that $\delta_D \equiv \delta_{K^{*-}K^+} -
\delta_{K^{*+}K^-} = 0^\circ~(180^\circ)$ indicates maximal
destructive (constructive) interference between the $K^*$
amplitudes. We consider thirteen resonant components (see
Table~\ref{tab:reslist}) as well as a uniform non-resonant
contribution. Dalitz plot analyses are only sensitive to relative
phases and amplitudes, hence we may arbitrarily define the amplitude
and phase for one of the two-body decay modes. The mode with the
largest rate, $K^{*+}K^{-}$, is assigned an amplitude $a_{K^{*+}K^-}
= 1$ and phase $\delta_{K^{*+}K^-} = 0^\circ$.

The efficiency for the selection requirements described above is not
expected to be uniform across the Dalitz plot because of the
momentum dependent reconstruction algorithms near the edge of phase
space. To study these variations, we produce Monte Carlo generated
$D^{*+} \rightarrow \pi^+_s D^0,~D^{0} \rightarrow K^{+} K^{-}
\pi^{0}$ events (based on {\footnotesize{GEANT3}} \cite{GEANT})
which uniformly populate the allowed phase space and pass them
through our event processing algorithms. We observe a modest and
smooth dependence of reconstruction efficiency on Dalitz plot
position and fit this to a two dimensional cubic polynomial in
($m^2_{{K^+}{\pi^0}},~m^2_{{K^+}{K^-}}$). The average reconstruction
efficiency for the decay chain $D^{*+}\rightarrow\pi^+_s D^0,~D^{0}
\rightarrow K^{+} K^{-} \pi^{0}$ in our signal region is found to be
$(5.8 \pm 0.1)\%$.

Figure~\ref{fig:FigureOne} shows that the background is significant.
To construct a model of the background shape, we consider events in
the data sideband $3 < {\Delta}M < 10~\operatorname{MeV}\!/c^2$
within the $m_{{K^+}{K^-}{\pi^0}}$ signal region defined above, as
shown in Fig.~\ref{fig:FigureOne}(b). There are 384 events in this
selection, about three times the amount of background we estimate in
the signal region. The background is dominated by random
combinations of unrelated tracks and showers. Although the
background includes $K^{*\pm}$ and $\phi$ mesons combined with
random tracks and/or showers, these events will not interfere with
each other or with resonances in the signal. The background shape is
well fitted by a two dimensional cubic polynomial in
($m^2_{{K^+}{\pi^0}},~m^2_{{K^+}{K^-}}$) with non-interfering terms
that represent $K^{*\pm}$ and $\phi$ mesons.

\begin{table}
  \begin{center}
\begin{tabular}{c|c|c}
\hline
\hline Resonance $r$& $m_r~(\operatorname{GeV}\!/c^2)$ & $\Gamma_r~(\operatorname{GeV}\!/c^2)$ \\
 \hline $K^*(892)^\pm$& $0.8917$& $0.0508$ \\
 $\phi(1020)$& $1.0190$& $0.0043$ \\
 non-resonant&
  $\text{flat}$&
  $\text{flat}$     \\
\hline
 $a_0(980)$& $0.9910$&   $0.0690$     \\
 $f_0(1370)$& $1.3500$&   $0.2650$     \\
 $K_0(1430)^\pm$& $1.4120$&   $0.2940$     \\
 $K_2(1430)^\pm$& $1.4260$&   $0.0985$     \\
 $f_0(1500)$& $1.5070$&   $0.1090$     \\
 $f'_2(1525)$& $1.5250$&   $0.0730$     \\
\hline
 $\kappa^\pm$& $0.8780$&   $0.4990$     \\
\hline \hline
\end{tabular}
  \end{center}
\caption{The masses and widths of resonances $r$ considered in this
analysis \protect\cite{PDG,a0Ref,f0Ref,KappaRef}.}
\label{tab:reslist}
\end{table}

\begin{table}
  \begin{center}
\begin{tabular}{lc|c|c}
\hline \hline & Amplitude& Phase $(^\circ)$&
Fit Fraction (\%)\\
\hline {Fit A} \\ \scriptsize{$\text{SL} \!= \!18.6\%$}\\
\hline $K^{*+}$& 1 (fixed)& $0$ (fixed)&
$46.1\pm3.1$   \\
 $K^{*-}$& $0.52 \pm 0.05 \pm 0.04$& $332 \pm 8 \pm 11$&
$12.3\pm2.2$    \\
 $\phi$&
   $0.64 \pm 0.04$&
    $326 \pm 9$  &
     $14.9\pm1.6$     \\
NR&
  $5.62 \pm 0.45$&
  $220 \pm 5$ &
   $36.0\pm3.7$     \\
\hline {Fit B}\\ \scriptsize{$\text{SL} \!= \!17.2\%$}\\\hline
 $K^{*+}$& 1 (fixed)& $0$ (fixed)&
$48.1\pm4.5$   \\
 $K^{*-}$& $0.52 \pm 0.05$& $313 \pm 9$&
$12.9\pm2.6$    \\
 $\phi$&
   $0.65 \pm 0.05$&
    $334 \pm 12$  &
     $16.1\pm1.9$     \\
$\kappa^+$&
  $1.78 \pm 0.43$&
  $109 \pm 17$ &
   $12.6\pm5.8$     \\
   $\kappa^-$&
  $1.60 \pm 0.29$&
  $128 \pm 17$ &
   $11.1\pm4.7$     \\

\hline \hline
\end{tabular}
  \end{center}

\caption{Dalitz plot fit results. The model for Fit A includes
$K^{*\pm}$, $\phi$, and a non-resonant contribution. The model for
Fit B includes $K^{*\pm}$, $\phi$, and $\kappa^\pm$. A significance
level ($\text{SL}$), calculated by the method of Ref.~\cite{sl}, is
shown for each fit.} \label{tab:fitresults}
\end{table}

We use the background and efficiency parameterizations in our Dalitz
plot fit to the data. Our results are presented in
Table~\ref{tab:fitresults}. Fit A includes the $K^*(892)^\pm$ and
$\phi(1020)$ resonances plus an interfering non-resonant (NR)
component and is shown in Fig.~\ref{fig:FigureTwo}(b)-(d). For each
entry in Table~\ref{tab:fitresults}, the first error shown is
statistical. Systematic errors are also shown for the $K^*$
submodes, since those are the results that ultimately contribute to
the phase difference and relative amplitudes this analysis seeks to
measure. The determination of these systematic errors is discussed
below.

Since it is difficult to distinguish a simple NR contribution from a
broad $S$-wave component, we investigated the effect of replacing
the NR component of Fit A with broad $S$-wave $\kappa^\pm
\rightarrow K^\pm\pi^0$ resonances parameterized using Breit-Wigner
amplitudes \cite{KappaRef}. The result of this substitution is shown
as Fit B in Table~\ref{tab:fitresults}. Both Fit A and Fit B have
good significance levels, and the projections of Fit A and Fit B are
indistinguishable, hence we have no reason to prefer one fit over the other. Significance levels are calculated by the method of Ref.~\cite{sl}.

We tested other combinations of broad amplitudes as possible
replacements to the simple non-resonant component, including one fit
with $K_0(1430)^\pm \rightarrow K^\pm\pi^0$ and $\kappa^\pm
\rightarrow K^\pm\pi^0$ and another fit with a NR component combined
with $K_0(1430)^\pm$. We did not find that either of these fits were
preferable to Fit A or Fit B, although we do include these results
when determining our model systematic error. We did not find
significant evidence for any of the other resonances listed in
Table~\ref{tab:reslist}. A fit which included only $K^{*\pm}$ and
$\phi$ contributions (without a NR component) was significantly
worse than Fit A.

Since the choice of normalization, phase convention, and amplitude
formalism may not always be identical for different experiments, fit
fractions are reported in addition to amplitudes. The fit fraction
is defined as the integral of a single component (resonant or
non-resonant) over the Dalitz plot, divided by the integral of the
coherent sum of all components over the Dalitz plot \cite{Tim}. The
sum of the fit fractions for all components will not necessarily be
unity because of interference in the coherent sum.

We use the full covariance matrix from Fit~A and Fit~B to determine the
statistical errors on the fit fractions and to properly include the
correlated components of the uncertainty on the amplitudes and
phases. After each fit, the covariance matrix and final parameter
values are used to generate a large number of sample parameter sets.
Fit fractions are calculated as described above for each set of
parameters, and the Gaussian widths of these distributions represent
the statistical errors on the nominal fit fractions.

The strong phase difference $\delta_D$ and relative amplitude $r_D$
are defined as follows:
\begin{eqnarray}
   {r}_{D} e^{i\delta_D} = \frac{a_{K^{*-}K^{+}}
}{a_{K^{*+}K^{-}}
}~e^{i(\delta_{K^{*-}K^{+}}-\delta_{K^{*+}K^{-}})} ,
\label{eqn:rDdeltaD}
\end{eqnarray}
where $r_D$ in Eq.~(\ref{eqn:rDdeltaD}) is defined as real and
positive. The strong phase difference is equivalent to the overall
phase difference due to our assumption that CP violation in $D$
decays is negligible. With this definition we can simply read our
nominal results from Fit A of Table~\ref{tab:fitresults}:
$$\delta_{D} = 332^\circ \pm 8^\circ \pm 11^\circ, ~~{r}_{D} = 0.52 \pm 0.05 \pm 0.04.$$

We consider systematic errors from experimental sources and from the
decay model separately. Contributions to the {\it experimental}
systematic uncertainties arise from our models of the background,
the efficiency, the signal fraction, and the event selection. Our
general procedure is to change some aspect of the analysis and
interpret the change in the values of the amplitude ratio $r_D$ and
phase difference $\delta_D$ as an estimate of the associated
systematic uncertainty. In Fit A, we fix the coefficients of the
background parameterization to the values found in our fit to the
sideband region as described above. To estimate the systematic
uncertainty on this background shape, we perform a fit where these
coefficients are allowed to float constrained by the covariance
matrix of the background fit. A similar method is used to determine
the systematic uncertainty for the efficiency shape. We change
selection criteria in the analysis to test whether our Monte Carlo
simulation properly models the efficiency. We vary the minimum
$\pi^0$ daughter energy, the cuts on $m_{K^+K^-\pi^0}$ and
${\Delta}M$, the $D^{*+}$ minimum momentum fraction, the
$m(\gamma\gamma) - m(\pi^0)$ requirement, and the RICH and specific
energy criteria. We allow the width of the $\phi(1020)$ to float to
accommodate detector resolution effects. We performed partial fits
of the Dalitz plot excluding regions not close to the $K^\ast(892)$
bands, and we changed the invariant mass-squared variables in our
fits from ($m^2_{{K^+}{\pi^0}},~m^2_{{K^+}{K^-}}$) to
($m^2_{{K^-}{\pi^0}},~m^2_{{K^+}{\pi^0}}$). The largest experimental
systematic uncertainties are $\pm8^\circ$ for $\delta_D$ when
allowing the background parameters to float, and $\pm0.05$ for $r_D$
when allowing the efficiency parameters to float, as described
above.

The {\it model} systematic error arises from uncertainty in the
choice of resonances used to fit the Dalitz plot. We fit the data to
many models that incorporate various combinations of the resonances
listed in the lower part of Table~\ref{tab:reslist} in addition to
the $K^*(892)^\pm$ and $\phi(1020)$. We allow our $\kappa$ mass and
width to float in a separate fit, finding the preferred values to be
$m_\kappa = (855 \pm 15) \operatorname{MeV}$ and $\Gamma_\kappa =
(251 \pm 48) \operatorname{MeV}$. The significance level for the fit
where the $\kappa$ mass and width float is 16.2\%. The floating
$\kappa$ mass is consistent with Ref.~\cite{KappaRef}, but the
floating $\kappa$ width is smaller by about two standard deviations.
We use our measured error and the error from Ref.~\cite{KappaRef} on
$\Gamma_\kappa$ to calculate the deviation.

We determine the total experimental and model systematic uncertainties
separately. We take the square root of the sample variance of the
amplitudes and phases from the nominal result compared to the
results in the series of fits described above as a measure of the
systematic uncertainty. We find $\pm 2.9^\circ (\operatorname{exp.})
\pm 10.6^\circ (\operatorname{model})$ for
$\delta_D$ and $\pm 0.016
(\operatorname{exp.}) \pm 0.038 (\operatorname{model})$ for $r_D$. Adding systematic
errors in this way results in a model systematic error for $\delta_D$ that is less than the
difference in $\delta_D$ when comparing Fit A to Fit B. We add the experimental and
model systematic uncertainty in quadrature to obtain the total
systematic uncertainty reported in Table~\ref{tab:fitresults}.

Our systematic error is dominated by the model dependence,
    and the largest deviations from the nominal fit were observed in the
series of fits where we replaced the non-resonant contribution with
the $\kappa^\pm$. If fits including a $\kappa^\pm$ resonance are
removed from consideration, then the systematic errors on $\delta_D$
and $r_D$ decrease from $\pm11^\circ$ and $\pm0.04$ to $\pm8^\circ$
and $\pm0.03$, respectively, and the remaining systematic
uncertainty is dominated by fits including the $K_0(1430)^\pm$.

As a cross-check, we estimate the branching ratio of $D^0
\rightarrow K^{+}K^-\pi^0$ from our data and compare it to the
published value. Branching ratio measurements are not the focus of
this analysis, so systematic errors have not been investigated.
Based on our $m_{K^+K^-\pi^0}$ fit, we have a total of $627 \pm 30$
signal events. We may estimate the total number of $D^0$s expected
from continuum $D^{*+}$s in our data sample, based on the integral
cross-section for continuum $D^{*+}$ production near $\sqrt{s} =
10.6$~GeV \cite{CharmFragmentation}: $\sigma(e^+ e^- \rightarrow
D^{*+} X) = (583 \pm 8 \pm 33 \pm 14)~\operatorname{pb}$, where the
fourth error stems from
 external branching fraction uncertainties.
 From this information and
$BR(D^{*+}\rightarrow\pi^+_s D^0)$ \cite{PDG}, we estimate $BR(D^0
\rightarrow K^{+}K^-\pi^0) = (0.30 \pm 0.02 )\%$, which is significantly higher than the previous measurement
\cite{PDG,oldKKpi}. Combining our fit fractions with known values for $BR(K^{*\pm} \rightarrow
K^\pm \pi^0)$ and $BR(\phi \rightarrow K^+ K^-)$ \cite{PDG}, we
also estimate branching ratios of the resonant decay modes. We find
$BR(D^{0} \rightarrow K^{*+}K^{-}) = (0.38 \pm 0.04 )\%$, $BR(D^{0}
\rightarrow K^{*-}K^{+}) = (0.10 \pm 0.02 )\%$, and $BR(D^{0}
\rightarrow \phi{\pi^0}) = (0.084 \pm 0.012 )\%$. These branching
ratios are consistent with published measurements
\cite{PDG,oldKstarCLEO,oldKstarARGUS,PhiPi0BELLE}.

$U$-spin symmetry \cite{Theory} predicts the following for $D^0$
decays to a pseudoscalar meson
 and a vector meson:
\begin{eqnarray}
{\cal{A}}(D^{0} \rightarrow \pi^{+} \rho^{-}) = -{\cal{A}}(D^{0}
\rightarrow K^{+} K^{*-})
 \label{eqn:piKplus}
\end{eqnarray}
and
\begin{eqnarray}
{\cal{A}}(D^{0} \rightarrow \pi^{-} \rho^{+}) = -{\cal{A}}(D^{0}
\rightarrow K^{-} K^{*+}) \label{eqn:piKminus},
\end{eqnarray}
where $\cal{A}$ is the respective dimensionless invariant amplitude
for each decay. Dividing Eq.~(\ref{eqn:piKplus}) by
Eq.~(\ref{eqn:piKminus}), and assuming that ${\cal{A}}(K^{*+}
\rightarrow K^{+} \pi^{0}) = {\cal{A}}(K^{*-} \rightarrow K^{-}
\pi^{0})$, gives:
\begin{eqnarray}
\frac{{\cal{A}}(D^{0} \rightarrow \pi^+\rho^- )}{{\cal{A}}(D^{0}
\rightarrow \pi^-\rho^+)} = \frac{a_{K^{*-}K^{+}}
}{a_{K^{*+}K^{-}}
}~e^{i(\delta_{K^{*-}K^{+}}-\delta_{K^{*+}K^{-}})}. \label{eqn:piKfrac}
\end{eqnarray}

Assuming a phase convention such that a phase difference of
$0^\circ$ indicates maximal destructive interference between $\rho^-$ and
$\rho^+$, and assuming ${\cal{A}}(\rho^{+} \rightarrow \pi^{+}
\pi^{0}) = {\cal{A}}(\rho^{-} \rightarrow \pi^{-} \pi^{0})$, we can
use the recently published results of a Dalitz plot analysis of
$D^0\rightarrow\pi^+\pi^-\pi^0$ \cite{A} to evaluate the left hand
side of Eq.~(\ref{eqn:piKfrac}):
$$(0.65\pm0.03\pm0.02)e^{i(356^\circ\pm3^\circ\pm2^\circ)}$$
which may be compared to the right hand side of
Eq.~(\ref{eqn:piKfrac}) which comes from this analysis:
$$(0.52\pm0.05\pm0.04)e^{i(332^\circ\pm8^\circ\pm11^\circ)}.$$

In conclusion, we have examined the resonant substructure of the
decay $D^{0} \rightarrow K^{+} K^{-} \pi^{0}$ using the Dalitz plot
analysis technique. We observe resonant $K^{*+}K^{-}$,
$K^{*-}K^{+}$, and $\phi\pi^0$ contributions.  We also observe a
significant $S$-wave modeled as a $\kappa^{\pm}K^\mp$ or a
non-resonant contribution. We determine $\delta_{D} = 332^{\circ}
\pm 8^{\circ}  \pm 11^{\circ}$ and $r_{D} = 0.52 \pm 0.05 \pm 0.04 $.

We gratefully acknowledge the effort of the CESR staff in providing
us with excellent luminosity and running conditions.
D.~Cronin-Hennessy and A.~Ryd thank the A.P.~Sloan Foundation. This
work was supported by the National Science Foundation, the U.S.
Department of Energy, and the Natural Sciences and Engineering
Research Council of Canada.

\end{document}